\documentclass[11pt]{article}
\usepackage{graphicx}
\usepackage{longtable,lscape}
\usepackage{amsthm}
\usepackage{amsfonts}
\usepackage{amsmath}
\usepackage{bbm}
\usepackage{float}
\usepackage{url}

\newtheorem{theorem}{Theorem}

\newcommand{\captionfonts}{\footnotesize}
\makeatletter  % Allow the use of @ in command names
\long\def\@makecaption#1#2{%
  \vskip\abovecaptionskip
  \sbox\@tempboxa{{\captionfonts #1: #2}}%
  \ifdim \wd\@tempboxa >\hsize
    {\captionfonts #1: #2\par}
  \else
    \hbox to\hsize{\hfil\box\@tempboxa\hfil}%
  \fi
  \vskip\belowcaptionskip}
\makeatother 
\begin{document}
\title{Meaning-focused and Quantum-inspired Information Retrieval}
\author{Diederik Aerts$^1$, Jan Broekaert$^1$, Sandro Sozzo$^1$ and Tomas Veloz$^{2,1}$ \vspace{0.5 cm} \\ 
        \normalsize\itshape
        $^1$ Center Leo Apostel for Interdisciplinary Studies \\
        \normalsize\itshape
        and, Department of Mathematics, Brussels Free University \\ 
        \normalsize\itshape
         Krijgskundestraat 33, 1160 Brussels, Belgium \\
        \normalsize
        E-Mails: \url{diraerts@vub.ac.be,jbroekae@vub.ac.be, ssozzo@vub.ac.be}
          \vspace{0.5 cm} \\ 
        \normalsize\itshape
        $^2$ Department of Mathematics, University of British Columbia \\
        \normalsize\itshape
         Kelowna, British Columbia, Canada \\
        \normalsize
        E-Mail: \url{tomas.veloz@ubc.ca} \\
              }
\date{}
\maketitle
\begin{abstract}
\noindent
In recent years, quantum-based methods have promisingly integrated the traditional procedures in information retrieval (IR) and natural language processing (NLP).  Inspired by our research on the identification and application of quantum structures in cognition, more specifically our work on the representation of concepts and their combinations, we put forward a `quantum meaning based' framework for structured query retrieval in text corpora and standardized testing corpora. This scheme for IR rests on considering as basic notions, (i) `entities of meaning', e.g., concepts and their combinations, and (ii) traces of such entities of meaning, which is how documents are considered in this approach. The meaning content of these `entities of meaning' is reconstructed by solving an `inverse problem' in the quantum formalism, consisting of reconstructing the full states of the entities of meaning from their collapsed states identified as traces in relevant documents. The advantages with respect to traditional approaches, such as Latent Semantic Analysis (LSA), are discussed by means of concrete examples.
\end{abstract}

\section{Introduction\label{intro}}
Since the appearance of {\it The Geometry of Information Retrieval} \cite{vr2004}, introducing a quantum structure approach to Information Retrieval (IR), Widdows and Peters \cite{wp2003}, using a quantum logical negation for a concrete search system, and Aerts and Czachor \cite{aertsczachor2004}, identifying quantum structure in semantic space theories, such as Latent Semantic Analysis (LSA) \cite{ddflh1990}, the employment of techniques and procedures induced from the mathematical formalisms of quantum physics -- Hilbert space, quantum logic and probability, non-commutative algebras, etc. -- in fields such as IR and natural language processing (NLP), has produced a number of new and interesting results \cite{w2006,m2008,lc2008,za2010,pflr2010}. The latter can be placed within a growing quantum structure research in cognitive domains \cite{aertsaerts1995,aertsgabora2005ab,bruzaetal2007,bruzaetal2008,aerts2009,bruzaetal2009,pb2009,k2010,songetal2011,bpft2011,bb2012,busemeyeretal2012}. These quantum-based approaches mainly integrate and generalize the standard procedures in IR and NLP. Roughly speaking, one considers `documents' and `terms' as basic ingredients, concentrating on the so-called `document-term matrix' which contains as entries the number of times that a specific term appears in a specific document. Both terms and documents are represented by vectors in a suitable (Euclidean) semantic space, and the scalar product between these vectors is a measure of the similarity of the corresponding documents and terms. This approach has extended to Latent Semantic Analysis (LSA) \cite{ddflh1990}, Hyperspace Analogue to Language (HAL) \cite{lb1996}, Probabilistic Latent Semantic Analysis (pLSA) \cite{h1999}, Latent Dirichlet Allocation (LDA) \cite{bnj2003}. 
Search engines on the World Wide Web, though introducing on top additional procedures, e.g., `page ranking', mostly rely on this linear space technique to determine a basic set of relevant documents. Nothwithstanding its success, the procedure meets some difficulties, including high computational costs and lack of incremental updates, which limits its applicability. Furthermore, one can claim the `ad hoc' character of the procedure and, as a consequence, none of the examples mentioned above is immune to criticisms.

Inspired by a two-decade research on the identification and application of quantum structures in disciplines different from the micro-world \cite{aertsaerts1995,aertsgabora2005ab,aerts2009,aerts1986,aerts1999b,as2011,abgv2012,ags2012,abgs2012}, we put forward in this paper the first steps leading to a possible conceptually new perspective for IR and NLP. In this approach, we want to replace terms by `entities of meaning' as primary notions, which can be concepts or combinations of concepts. Such `entities of meaning' can be in different states and change under the influence of the `meaning landscape', or `conceptual landscape', or `conceptual context'. If we say `pour out the water', and the meaning landscape is that of a flooded village after a heavy storm, the state of the entity of meaning which is `pour out the water' is very different, from when the meaning landscape is that of a cafe where we are having some refreshments together. The explicit act of considering entities of meaning in states makes our approach fundamentally `contextual'. Moreover, documents are not regarded as collection of words, but as traces, i.e. more concrete states, of these entities of meaning, or concepts, or combinations of concepts. This means that a document is considered to be a collapse of full states of different entities of meaning, each entity leaving a trace in the document. The words are only spots of these traces and they are not the main meaning carriers. The technical focus of our approach consists in trying to reconstruct there full states of the different entities of meaning from experiments that can only spot their traces, i.e. that can only look at words in documents. We believe that aspects of our quantum approach to cognition, still in full development, can help in formulating and making technically operational this `inverse problem', consisting in `reconstructing the the full states of the different entities of meaning', starting from their collapsed states as traces of word spots in documents.

We introduce in Sec. \ref{novelties} the basic notions that are needed in our scheme for IR, that is, entities of meaning, which can be concept combinations, documents as traces of such entities of meaning and their technical reconstruction. We point out how our perspective is different from traditional approaches, e.g., LSA. We specify the sense in which the new paradigm is meaning focused and argue that it is closer to the processes concretely working in the human mind. Moreover, the quantum-theoretic formalism we have recently developed to model concept combinations is a possible natural candidate to represent our meaning-based scheme. Indeed, the pair (entities of meaning, documents) is replaced by the pair (concepts/combinations, exemplars) in Sec. \ref{brussels}. Thus, the quantum modeling we have employed in the simple case where entities of meaning are concepts and documents are exemplars can be naturally used also in these more general IR cases. We stress that we have as yet no theory -- but specific cases -- to solve the functional inverse problem in an IR system, hence we only sketch the first steps for an approach on the theoretical level. Nevertheless, our quantum-inspired scheme is potentially more performant than LSA-based techniques. This is explicitly shown in Sec. \ref{lsa}, where a LSA analysis of Hampton's data on disjunctions of concepts is supplied and compared with our quantum cognition model in Sec. \ref{brussels}. We draw the conclusion that (i) LSA is only partially capable of capturing the effects of overextension and underextension in concept combinations, (ii) LSA supplies only approximate solutions, unlike our quantum modeling approach. This suggests that an application of our quantum and meaning-based approach would be more efficient than classical approaches both in text analysis and in information recovering.

\section{Fundamentals of the meaning-based approach\label{novelties}}
We present here the basics of our meaning-based approach for IR, explaining, in particular, its novelties with respect to the traditional IR and NLP procedures and justifying the use of the quantum-mathematical formalism in Sec. \ref{brussels} in it. 

LSA and its extensions typically use word-counting techniques in which the semantic structure of large bodies of text is incorporated into semantic linear spaces and the `document-text matrix'. The latter contains as entries the number of times that a given term appears in a given document. If one labels the rows of this matrix by the documents and the columns by the terms, then each row can be viewed as a vector representing the corresponding document and each column as a vector representing the corresponding term. If vectors are normalized, their scalar product is a measure of the `similarity' of the corresponding documents and terms, hence this data analysis can be used in IR and NLP. In the approach we put forward in this paper, instead, the semantic structure of texts is incorporated directly into concepts and their traces describing documents. This method is closer to the processes concretely working in the human mind, which explains why the quantum modeling in Sec. \ref{brussels}, which faithfully describes human collected data on concept combinations, can be applied to IR in a straightforward way. We stress, however, that the approach we propose is not yet worked out sufficiently to be applied to concrete IR problems, since many of the technical aspects of the inverse problem need to be specified and elaborated in different ways. Notwithstanding this, we will see that some interesting conclusions can already be drawn.

The first fundamental element of our approach is the conceptually new fact that `terms' are replaced by `entities of meaning' as basic elements, and specifically such entities of meaning can be `concepts' or their combinations. This is why we work with `entities of meaning', usually expressed as concept combinations, rather than with terms. Our procedure takes into account the meaning of the words from the very beginning, and there are valuable reasons to believe that this is how the human mind works. Indeed, whenever we read a text, we understand the `meaning' of the text, and even have no efficient memory for the `structure of the terms'. The substitution of terms with entities of meaning also allows us to use the quantum modeling formalism in Sec. \ref{brussels} for representing concepts and their combinations. In the latter work, concepts are indeed treated as entities that can be in different states and change their state under the influence of a context, exactly as microscopic quantum particles change their state under a measurement context.

The second basic element is the interpretation of `pieces of texts' or, better, `documents', as `traces of these entities of meaning'. In this perspective, a document is not regarded as a combination of words but, rather, as  
collapsed states of the considered entities of meaning, or combinations of concepts. And, again, there are valuable reasons to believe that this view is an adequae representation of how the human mind operates. Indeed, if we consider the entity of meaning {\it The Cat Runs Through The Garden}, which is a combination of concepts, and consider a document telling about the adventures of a cat, a trace of {\it The Cat Runs Through The Garden} can be identified in this document, depending on the meaning content of the story about the cat. A weight can be identified representing the `aboutness' of the meaning content of the document with respect to the entity of meaning {\it The Cat Runs Through The Garden}. And the document itself can be considered as a collapsed state of the 
full state of the entity of meaning {\it The Cat Runs Through The Garden}. This is exactly the structure that we can study by means of the quantum modeling formalism, for example what we explained in Sec. \ref{brussels}, namely, a document being a collapsed state of an entity of meaning after a measurement process (a cognitive test on subjects, a query on the web, etc.). The preceding insight has been inspired by what typically occurs in quantum experiments on microscopic particles, where one looks for traces of quantum particles. Whenever an experimental test is performed, a trace, or snapshot, of a quantum particle is left in a suitable apparatus, e.g., a Bell chamber. A trace of this kind reveals a collapsed process of the quantum particle in the real physical space.

The third element of our approach is developing a technique to reconstruct the full states of the considered entities of meaning starting from weights that these full states contain with respect to their collapsed states, i.e. the documents. This is what in physics is called the `inverse problem'. We observe that the human mind 
performs this inverse problem brilliantly, it indeed reconstructs the `entities of meaning' of a document starting from what is written on the piece of paper, hence starting from -- not the words -- but the trace of these entities of meaning, or the collapsed states. And this is exactly also what quantum experimentalists and phenomenologists do, they recover the initial state of quantum particles starting from their collapsed states and outcome statistics of repeated experiments. 

We stress that the inverse problem above can be technically very complicated, so that we have only given a conceptual description of it here. One could, for example, investigate the methods employed in quantum physics for state reconstruction and tomography, extending them to IR systems. In any case, our quantum cognition approach has already performed a complete reconstruction of the inverse problem in a Hampton's test of typicality \cite{ags2012,abgs2012}. Test subjects were asked to choose from a list of 24 exemplars the one that they estimated best represented the concepts {\it Fruits}, {\it Vegetables} and their disjunction {\it Fruits or Vegetables}. We elaborated a 25-dimensional complex Hilbert space which perfectly agreed with empirical data and allowed us to reconstruct and represent the initial states of the concepts. In this quantum model, the given concepts are the `entities of meaning', while the exemplars, being more concrete states, or traces, of concepts in our approach, play the role of collapsed states of these conceptual entities of meaning. This suggests that a similar Hilbert space scheme can be envisaged where we replace `exemplars' by `paragraphs of texts' playing the role of documents, while `concepts' are replaced by `entities of meaning'. Of course, a quantum-mechanical model of this kind needs to be specified once a real experiment is performed on human subjects, but it already contains the genuine quantum structures that play a role in an IR process, such as collapse, contextuality, emergence, entanglement, interference and superposition.

\section{Effectiveness of a quantum cognition modeling in IR\label{brussels}}
It is well known that classical logical and probabilistic approaches fail when dealing with conceptual vagueness, the gradation of membership weights and concept combination
(see, e.g., \cite{oshersonsmith1981,hampton1988a,hampton1988b,Hampton2007}).
For this reason, we have recently worked out a quantum-theoretic approach for concept combination \cite{aerts2009,as2011,abgv2012,ags2012,abgs2012}. On the other hand, we have anticipated in Sec. \ref{novelties} that this quantum cognition formalism is a natural candidate to represent our meaning based approach for IR. The first reason is that both approaches deal with concepts and their states and meaning, the second is that the meaning based approach for IR in Sec. \ref{novelties} rests on some processes that are hypothesized to work in the human mind, while our quantum-theoretic modeling faithfully accords with a large collection of experimental data on human subjects on the combination of two concepts \cite{hampton1988a,hampton1988b}. But there is a third and even stronger motivation for concretely using our quantum cognition approach in dealing with IR problems: it mathematically follows the same scheme of Sec. \ref{novelties}. Here, the role of `entities of meaning' is played by the concepts and their disjunctions/conjunctions, while the role of `documents' is played by the set of `exemplars'. Indeed, the latter are more concrete states of concepts, hence they can be regarded as traces, or collapsed states, of these conceptual entities of meaning. This means that our quantum modeling scheme would work `equally well' if we did the experiment with `concepts, combinations of concepts' and `real documents', with human subjects estimating the `aboutness' of certain concepts with respect to a document. It is thus worth focusing on this quantum cognition approach and compare it with traditional approaches.

To model combinations of two concepts we need a Fock space $\cal F$ which consists of two sectors: `sector 1' is a Hilbert space $\cal H$, while `sector 2' is a tensor product Hilbert space $\cal H \otimes \cal H$, so that ${\cal F}={\cal H} \oplus ({\cal H}\otimes {\cal H})$. As a general consideration, sector 1 mainly enables modeling of interference connected phenomena, while sector 2 mainly enables modeling of entanglement connected phenomena. Let us consider the membership weights of exemplars of concepts and their conjunctions/disjunctions measured by Hampton \cite{hampton1988a,hampton1988b}. He identified systematic deviations from classical (fuzzy set) conjunctions/disjunctions, an effect known as `overextension' or `underextension'. We concentrate on disjunctions here, which we will actually compare with LSA in Sec. \ref{lsa}. A completely similar analysis can be done for conjunctions \cite{aerts2009}. It can be shown that a large part of Hampton's data cannot be modeled in a classical probability space satisfying Kolmogorov's axioms, due to the following theorem.
\begin{theorem} \label{th3}
The membership weights $\mu_x(A), \mu_x(B)$ and $\mu_x(A\ {\rm or}\ B)$ of an exemplar $x$ for the concepts $A$, $B$ and $A \ {\rm or} \ B$ can be represented in a classical probability model if and only if the following two conditions are satisfied. 
\begin{eqnarray} \label{maxdeviation}
\Delta_d=\max(\mu_x(A),\mu_x(B))-\mu_x(A\ {\rm or}\ B)\le 0 \\ \label{kolmogorovianfactordisjunction}
0 \le k_d=\mu_x(A)+\mu_x(B)-\mu_x(A\ {\rm or}\ B)
\end{eqnarray}
where $\Delta_d$ is the \emph{disjunction maximum rule deviation}, and $k_d$ is the \emph{Kolmogorovian disjunction factor}.
\begin{proof} See (Aerts, 2009a), theorem 6.
\end{proof}
\end{theorem}
Equation (\ref{maxdeviation}) expresses compatibility with the maximum rule for the conjunction of fuzzy set theory and, more generally, with monotonicity of classical Kolmogorovian probability. A situation with $\Delta_d >0$ is called `underextension' \cite{hampton1988b}. Equation (\ref{kolmogorovianfactordisjunction}) expresses instead compatibility with additivity of classical Kolmogorovian probability. Equations (\ref{maxdeviation}) and (\ref{kolmogorovianfactordisjunction}) together provide necessary and sufficient conditions to describe the experimental membership weights $\mu_x(A), \mu_x(B)$ and $\mu_x(A\ {\rm or}\ B)$ in a Kolmogorovian probability space $(\Omega, \sigma(\Omega), P)$ ($\sigma(\Omega)$. In this case, indeed, events $P_A,P_B\in \sigma(\Omega)$ exist such that $P(E_A)=\mu_x(A)$, $P(E_B)=\mu_x(B)$, and $P(E_A \cup E_B)=\mu_x(A \ {\rm or} \ B)$. 

Let us consider a specific example. Hampton estimated the membership weight of {\it Donkey} with respect to the concepts {\it Pet}, {\it Farmyard Animal} and their disjunction {\it Pet or Farmyard Animal} finding the values $\mu_{Donkey}(Pet)=0.5$, $\mu_{Donkey}(Farmyard \ Animal)=0.9$, $\mu_{Donkey}(Pet \ {\rm or}\ Farmyard \ Animal)=0.7$. Thus, the exemplar \emph{Donkey} presents underextension with respect to the disjunction \emph{Pet or Farmyard Animal} of the concepts \emph{Pet} and \emph{Farmyard Animal}. We have in this case $\Delta_d=0.2\not\le0$, hence no classical probability representation exists for these data, because of Th. \ref{th3}. It can instead be proved that a quantum probability model in Fock space exists for these Hampton's data \cite{hampton1988b}.
\begin{theorem} \label{th4}
The membership weights $\mu_x(A), \mu_x(B)$ and $\mu_x(A\ {\rm or}\ B)$ of an exemplar $x$ for the concepts $A$, $B$ and $A \ {\rm or} \ B$ can be represented in a quantum probability model where
\begin{equation} \label{muAorB}
\mu_x(A\ {\rm or}\ B)=m_x^2(\mu_x(A)+\mu_x(B)-\mu_x(A)\mu_x(B))+n_x^2({\mu_x(A)+\mu_x(B) \over 2}+Int_{x}(A,B))
\end{equation}
where the numbers $m_x^2$ and $n_x^2$ are convex coefficients, i.e. $0 \le m_x^2, n_x^2 \le 1$, $m_x^2+n_x^2=1$, and ${\theta_x}$ is the \emph{interference angle} with
\begin{equation} \label{interferenceterm}
Int_{x}(A,B)=\sqrt{1-\mu_x(A)}\sqrt{1-\mu_x(B)}\cos\theta_x
\end{equation}
\begin{proof} See Aerts, 2009a \cite{aerts2009}.
\end{proof}
\end{theorem}
The term $\mu_x(A)+\mu_x(B)-\mu_x(A)\mu_x(B)$ is what one would expect for the disjunction in the case of classical probability. The term $Int_{x}(A,B))$ is instead the quantum interference term and it is responsible, together with the average ${\mu_x(A)+\mu_x(B) \over 2}$, of the deviations from classical expectations. The coefficients $m_x^2$ and $n_x^2$ measure the weights of sectors 2 and 1, respectively, of ${\cal F}$. For example, in the case of {\it Donkey} with respect to {\it Pet}, {\it Farmyard Animal} and {\it Pet or Farmyard Animal}, we have that Th. \ref{th4} is satisfied with $m_{Donkey}^2=0.26$, $n_{Donkey}^2=0.74$ and $\theta_{Donkey}=77.34^{\circ}$. 

Theorem \ref{th4} and its corresponding theorem for conjunctions -- which we do not report, for the sake of brevity -- contain the quantum probabilistic expressions allowing the modeling of a large amount of Hampton's data \cite{hampton1988a,hampton1988b}. In particular, the quantum modeling above perfectly agrees with the data reported in Sec. \ref{lsa} and compared with LSA data. We have also proposed an explanation for the fact that a quantum approach of this kind is so successful in modeling the large collection of data by Hampton. We have hypothesized a mechanism in which a genuine quantum effect comes into play, namely `emergence'. Two processes, a logical one and a conceptual one, occur simultaneously in the human mind, and our quantum approach in Fock space enables both processes to be modeled.

We have seen above the deep reasons why our quantum cognition modeling can be successfully applied to IR. In Sec. \ref{lsa}, we will compare LSA and the quantum model modeling Hampton's data on concept combinations~\cite{hampton1988a,hampton1988b}. To conclude, we remark that, though our approach is conceptually different form standard IR approaches, such as LSA, it rests on similar basic ideas, that is, `meaning is expressed in texts by the environment of a term'. We are convinced however that coherence, emergence and contextuality, and their quantum modeling can express meaning in a way that is similar to how meaning is captured by the human mind.

\section{A comparison with LSA\label{lsa}}
Before analyzing Hampton's data by means of LSA, it is worth dwelling on two aspects that allow one to better grasp the connections between LSA and our quantum cognition modeling in Sec. \ref{brussels}.
 
(i) LSA typically calculates `similarity' through a complex technical procedure, which involves a real linear space representation of terms, a document-term matrix, a rank lowering through the reduction to a diagional matrix by singular value decomposition, dropping small eigenvalues, de-noisifying, and finally introducing values different from zero also for terms that do not appear in documents. In this way, one captures the `latent nature of similarity'. The calculation results, though most probably correlated to `similarity when tested on human subjects', do not express the latter directly. This entails that a LSA analysis of membership weights data, as well as a model for membership weights based on similarity \cite{Hampton2007}, do make sense in this case.

(ii) A LSA process introduces cuts and approximations, so that it models experimental data only approximately. This is usually considered an advantage, at least with respect to an approach, like our quantum-theoretic modeling in Secs. \ref{brussels} and \ref{novelties}, which can deliver models that `fit data completely'. Indeed, since one usually maintains that experiments are not perfect, one is led to believe that approaches that model these data approximately have a bigger chance to be close to reality, than approaches that model these data perfectly. There is en error in the above reasoning linked to the difference between `models that derive from a theory' and `ad hoc models'. An ad hoc model is a model specifically made for a situation, and for such a model it would indeed be suspicious if it could fit data correctly. A model that derives from a theory, when fitting data correctly, does not constitute a problem. Indeed when for such a model slightly different data are to be fit, this is also possible by varying some of the parameters. The latter remark expresses a fundamental difference between our quantum modeling and LSA. In our modeling, `also data that would be slightly different can again be perfectly fitted', which indicates that our models derive from a theory, i.e. quantum theory. This is `not' true for LSA: there is `no' corpus of texts that would fit, e.g., Hampton's data, as we see in what follows, which indicates that LSA is closer to an ad hoc way of model building.

We considered Hampton's membership weight data for the disjunction of eight pairs of concepts and 25 exemplars for each pair \cite{aerts2009,hampton1988b}. We computed the similarity between the exemplars of the concepts, and the similarity with respect to their disjunction, using the LSA Colorado website.\footnote{See the link {\it http://lsa.colorado.edu/}.} The LSA similarity website could not compute the similarity of one exemplar in three pairs of concepts, and of two exemplars on one pair of concepts. Therefore, we computed the similarity of 187 exemplars within the eight pairs of concepts in total.

The aim of this analysis is twofold. Firstly, we test whether the LSA similarity between the exemplar of a concept and the term denoting the concept can be used to estimate Hampton's membership data. To this end, we compare the LSA similarity with Hampton's membership data, and we also verify whether or not a membership model based on similarity, the `threshold model' \cite{Hampton2007}, improves the LSA estimations. The threshold model is a simple model which assigns membership weight zero to the exemplars that are below a similarity threshold $s_l$, assigns membership weight one to the exemplars that are above a similarity threshold $s_h$, and using a parameter $s_t$ builds a quadratic function to assign membership in the range $[s_l,s_h]$. Secondly, we identify the type of data that the LSA similarity (and the threshold model) delivers. To this aim, we compare the average number of exemplars that verify (or do not verify) Eqs. (\ref{maxdeviation}) and (\ref{kolmogorovianfactordisjunction}) in Hampton's data, in the LSA data, and in the threshold model. Whenever Eqs. (\ref{maxdeviation}) and (\ref{kolmogorovianfactordisjunction}) are satisfied, we call the exemplar of a `classical' type. If Eq. (\ref{maxdeviation}) is not satisfied, we call the exemplar of a $\Delta_d$ type, and if Eq. (\ref{kolmogorovianfactordisjunction}) is not satisfied, we call the exemplar of a $k_d$ type. Note that both inequalities cannot be violated simultaneously, and when one of them is violated, the data cannot be modeled by a Kolmogorovian probability model (see Th. \ref{th3}). 

Note that the LSA-similarity function can be negative. However, the membership function is assumed to be non-negative. We therefore  set the negative similarities to be equal to zero. Indeed, this is not a significant modification to our data set because only 10 of the 187 tested instances deliver negative similarities, and none of these values are lower than $-0.1$. 

\begin{figure}[h!]
\begin{center}
\includegraphics[height=8cm,width=17cm]{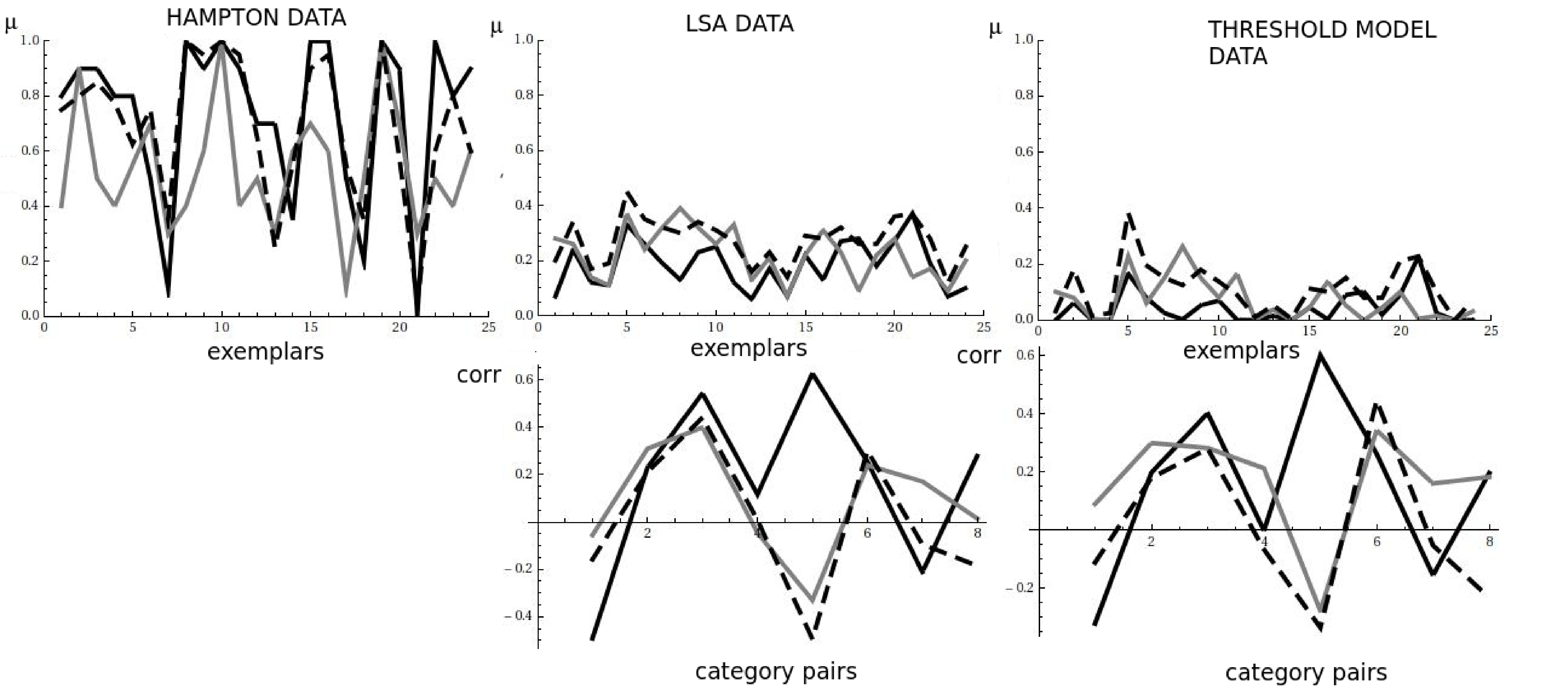}
\caption{Contrasting Hampton's data to LSA and Threshold models.} 
\label{data-analysis}	
\end{center}
\end{figure}

\begin{figure}[h!]
\begin{center}
\includegraphics[height=5cm,width=15cm]{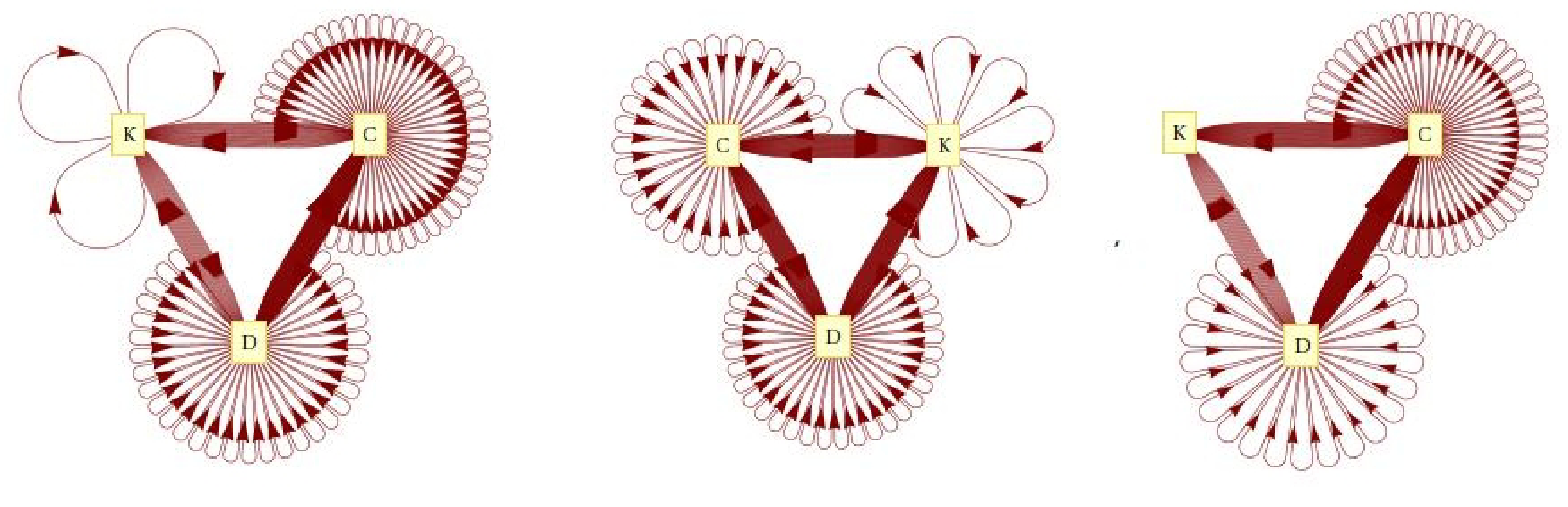}
\caption{Contrasting Hampton's data-type to LSA and Threshold models.} 
\label{graphs}	
\end{center}
\end{figure}

For reasons of space, we cannot compare concept by concept the performance of LSA and the threshold model in fitting Hampton's data. However, we illustrate that neither approach performs well. In Fig. \ref{data-analysis}, we compare the LSA and threshold models, using $s_l=0.1,s_t=0.5,$ and $s_h=0.9$,
 to the Hampton's data. The top row shows from left to right Hampton's data, the LSA similarity, and the threshold model data for the concepts {\it Home Furnishing} in black, {\it Furniture} in grey, and {\it Home Furnishing or Furniture} in dashed black. The exemplars of the concepts are on the x-axis, and the membership weights on the y-axis. It is clear that the range of values that both LSA and threshold model deliver are not close to the actual membership weigths measured by Hampton. The second row shows the Pearson correlation between Hampton's data and the LSA model in the center, and the threshold model on the right. The x-axis identifies the concept pair, and the y-axis identifies the correlation found for each concept, and their combination, with respect to Hampton's data. The coloring of the curves is the same as in the first row: the first concept of the pair is plotted in black, the second concept in grey and their combination in dashed black. We see that there is not significant correlation for any concept in both models. We therefore conclude that LSA and the threshold model using LSA data deliver weak estimations of Hampton's data, in their values and co-variations (correlations).
To compare the different types of Hampton's data with the LSA and threshold models, we compute a graph as follows: we define three nodes $\{C,D,K\}$, referring to the types 
`classical', $\Delta_d$, and $k_d$, respectively, and we build an edge $x\to y$ for each exemplar of each pair of concepts. The edge $x\to y$ indicates that the exemplar is of the type $x$ for Hampton's data, while it is of the type $y$ for the model we consider. For example, in the LSA data graph, if an arrow is such that $D\to C$, it means that one exemplar is of the type $\Delta_d$ for Hampton's data and of the type `classical' for the LSA data. Self-loops indicate no type difference between the two data sets, and so on. We draw these graphs in Fig. \ref{graphs}. The LSA data graph is shown on the left, the threshold model data, with parameters $s_l=0.1, s_t=0.5$ and $s_h=0.9$, is shown in the center, and  the threshold model with parameters $s_l=0.3, s_t=0.5$ and $s_h=0.7$, is shown on the right. Note that we consider all the exemplars of all concept pairs in this graph. In this sense we plot the average behavior of the models within the concepts. We do not show the detailed analysis for each pair of concepts, for reasons of space, but we mention that the tendencies we observe in the average case are also strong in the majority of the concept pairs.
For the three models, the tendency of having the same type for a given exemplar follows the order $C,D,K$. Where $C$ and $D$ are much larger than $K$. Moreover, increasing the threshold region, i.e. increasing $s_l$ and decreasing $s_h$, we observe that $C$ becomes even larger than $D$, and $K$ decreases to zero.
In addition, there is a strong tendency for $D$ exemplars in Hampton's data to become $C$ elements in the other models. As to the other transitions, there is not a concrete preference for the transition $D\to K$ and $K\to D$, except for the right plot, where we observe a larger amount of transitions going from $D$ to $K$ than from $K$ to $D$. Moreover, there is no preference for the transition $C \to K$ and $K \to C$. Indeed we count a similar amount of cases for each transition. We infer that LSA and the threshold model using LSA similarity have a weak capacity to identify instances of the $\Delta_d$ type, and that they cannot discern properly among instances of the classical and $k_d$ type. However, they have a non-neglectible capacity to identify classical data.

We can draw some general conclusions from the above analysis, as follows.

(i) The `bag of words' way of functioning of LSA, where terms and documents are not `entities of meaning' and `collapses of these entities of meaning', does not give rise to data that are in good agreement with what the human mind does in the same situation.

(ii) Notwithstanding (i), LSA captures quite some of the non-classical aspects of underextension and overextension. This can be technically understood from the point of view of our Fock space modeling in Sec. \ref{brussels}, as follows.

(ii.a) Word vectors are summed, which implies that LSA mainly works in sector 1 of our Fock space model. Actually, semantic spaces are real rather than complex linear spaces, but this is enough to introduce the possibility of describing some kind of interference. On the other hand, since `only' sector 1 of Fock space is taken into account in LSA, the bag of words problem is present: no difference can be made between `John hits Mary' and `Mary hits John'. This ordering problem is avoided in our Fock space approach by also taking into account sector 2, which is a tensor product.

(ii.b) The exclusion in LSA of the smallest eigenvalues after diagonalization introduces `latent values', i.e. weights become different from zero between terms and documents even if a term does not appear in a document. By means of this simplification technique, LSA manages to grasp something that is closer to the `states of the entities of meaning'. But it does so in a completely not understood way, as a by-product of a technique. This suggests that it should be possible to find more sophisticated techniques that directly and openly work toward the construction of `states of the entities of meaning'. This is exactly what the meaning-based and quantum-inspired approach we put forward in the present paper aims at. And, more, this is done in a way that we believe to be 
similar to what the human mind does.

\end{document}